Superconducting behavior of the solid solution, $Y_2Pd_{1-x}Pt_xGe_3$

Kartik K Iyer and E.V. Sampathkumaran[*]
Tata Institute of Fundamental Research, Homi Bhabha Road, Colaba, Mumbai – 400005, India.

**Abstract**

The compound $Y_2PdGe_3$ was earlier reported by us to be one of the very few ternary superconducting compounds ($T_c$= 3 K) belonging to the same structure as that of $MgB_2$. Here we report the results of electrical resistivity, magnetization and heat capacity measurements at low temperatures on the solid solution with a nominal starting composition, $Y_2Pd_{1-x}Pt_xGe_3$, to understand the influence of gradual replacement of Pd by Pt on $T_c$. The superconducting properties of this solution is distinctly interesting in the sense that the $T_c$ varies *monotonically* with increasing x in sharp contrast to the *non-monotonic* variation for other isostructural solid solutions reported recently.

Subsequent to the discovery of superconductivity in $MgB_2$ [1] below 39 K, the compound, $Y_2PdGe_3$, was reported by us [2] to be the first superconducting ternary intermetallic compound with the same crystal structure ($AlB_2$-derived hexagonal, space group P6/mmm), with a transition temperature ($T_c$) of 3 K. Band structure calculations suggest [3] that the Y4d states dominating at the Fermi level ($E_F$) are presumably responsible for superconductivity, while, in the case of $MgB_2$, the valence electrons from the B-site have been believed to play a major role. Subsequently, few more superconducting ternary and quasi-ternary compounds with this structure were identified: $Sr(Ga_{0.37}Si_{0.63})_2$ [Ref. 4], $Ca(Al_{0.5}Si_{0.5})_2$ [Ref. 5, 6], $Ca(Al_{1-x}Ga_x)Si$ [Ref. 7], $CaCu_{2-x}Si_x$ [Ref. 8], $YPt_{0.5}Ge_{1.5}$ [Ref. 9], and $Y_2PdGe_{3-x}Si_x$ [Ref. 10]. It was generally found that there is a non-monotonic variation of $T_c$ with $x$ - irrespective of which element is substituted and of the nature of the substituting element - in all these solid solutions. This is true even for an isoelectronic substitution at the Ge site in $Y_2PdGe_3$, that is, in $Y_2PdGe_{3-x}Si_x$; that is, $T_c$ initially increases (in the range x= 0.4) and then falls, whereas the lattice constant $a$ ($c$) decreases (increases) obeying Vegard's law. In order to explore whether this trend is applicable to an isoelectronic substitution at the Pd site, it is of interest to explore how $T_c$ varies in the series, $Y_2Pd_{1-x}Pt_xGe_3$. We have therefore attempted to synthesize this solid solution for the first time and investigated its low temperature behavior by dc electrical resistivity ($\rho$), dc magnetic susceptibility ($\chi$) and heat capacity (C) studies. The results reveal that, interestingly, there is a monotonic variation of $T_c$ with $x$ in this solid solution.

All the samples, $Y_2Pd_{1-x}Pt_xGe_3$ (x=0.0, 0.1, 0.2, 0.3, 0.4, 0.5, 0.7 and 1.0), were prepared by melting together stoichiometric amounts of constituent elements (purity >99.9%) in an arc furnace in an atmosphere of high-purity argon. All the ingots were then sealed in a quartz tube and subsequently annealed at 1123 K for a week. The samples were characterized by x-ray diffraction (Cu $K_\alpha$) and the patterns (see figure 1) indicated the presence of a weak secondary phase with $ThCr_2Si_2$-type tetragonal structure as known earlier [2,9,11], in addition to the main $AlB_2$-type hexagonal phase. Since such a tetragonal phase containing Y, Pt, Pd and Ge have not been known to be non-superconducting, the presence of this weak extra phase does not interfere with our conclusions. We have also looked at all the samples by scanning electron microscope and the parasitic phase could be detected. The composition of the main phase obtained by energy dispersive x-ray (EDX) analysis is typically $Y_2(Pd,Pt)_{1.3}Ge_{2.5}$ for the entire solid solution (including $x$= 0.0), in good agreement with Ref. 9 for $x$= 1.0. Therefore, the properties described in the rest of the article correspond to this composition, though nominal composition only is mentioned throughout this text. At this juncture, it may be recalled that the defect $AlB_2$ structure with a vacancy ordering (at the Ge-site) has been known for $YGe_{2-x}$ compounds [12]. The composition of the parasitic $ThCr_2Si_2$ phase was found to be significantly deficient in Pd/Pt (e.g., $YPt_{0.25}Ge_2$) by EDX analysis. $\rho$ as a function of temperature (T) was measured (in the range 1.8K to 80K) with a four probe technique employing a commercial physical property measurements system (Quantum Design, USA). The same instrument was used to measure heat-capacity by relaxation method. The $\chi$ data in the presence of 50 Oe were taken below 10 K for the zero-field-cooled (ZFC) conditions of the specimens in the ingot form employing a commercial (Quantum Design, USA) superconducting quantum interference device (SQUID). In addition, we have performed measurements for ZFC and field-cooled (FC) conditions of the specimens in the powder form for x= 0.0, 0.5 and 1.0 with the help of a commercial (Oxford Instruments, UK) vibrating sample magnetometer (VSM) in the presence of 20 Oe.

The lattice constants (± 0.005 Å) derived, say from (422) and (224) lines, are plotted in figure 2 for the main phase ($AlB_2$-type). If there is an ordering of Pt/Pd and Si sites, then one observes superstructure lines. These lines are usually weak, often escaping detection. For this reason, in this article, we stick to primitive $AlB_2$ unit-cell, which means that $a$ and $c$ are not doubled unlike in Ref. 2 and 3. It is clear that $a$ decreases, whereas $c$ increases with x, in such a way that the unit-cell volume remains nearly constant within 1 Å$^3$. The x-dependence of $a$ and $c$ is somewhat similar to that noted for Si substitution for Ge [10], despite the fact that a 4d transition metal ion is being replaced by a 5d transition metal ion. This implies that there are significant changes in the d-orbital hybridization effects with the gradual replacement of Pd by Pt.

We show the ρ(T) behavior at low temperatures in figure 3. The residual resistivity values, typically given by that at 3.5 K, is nearly the same for the two end members (of the order of 240 μΩ cm). The values apparently are marginally increased for intermediate compositions, presumably due to chemical disorder induced by substitution. As reported earlier [2], for x= 0.0, the ρ starts falling at ($T_c^{onset}=$ ) 3.2 K and zero resistance is attained below ($T_c^0=$ ) 2.7 K. The width of the transition is thus nearly 0.5 K. As $x$ is increased, $T_c^{onset}$ as well as $T_c^0$ decrease monotonically with increasing Pt content, as shown in figure 4 (which includes $T_c^{midpoint}$ also), with $Y_2PtGe_3$ attaining a value of about 2.15 K. The width of the transition keeps decreasing with increasing substitution and this is an interesting observation considering that increasing chemical disorder should result in an opposite effect.

At this juncture, it may be recalled that a binary phase $YGe_{1.62}$, crystallizing in α-$ThSi_2$-type tetragonal structure, also has been known to superconduct below 2.4 K [13]. In view of this, though *we did not find any extra lines attributable to this phase in the x-ray diffraction pattern,* it is important to ensure that superconductivity as detected in the ρ data does not arise from such an impurity phase, particularly noting that our value of $T_c$ is different from that reported for arc-melted $Y_2PtGe_3$ in Ref. 9. We have therefore measured heat capacity in the temperature region of interest. We observe a prominent jump in C(T) at the respective transition temperatures for all compositions (Fig. 5) and the observed magnitudes of the jump at $T_c$ for all compositions are comparable thereby establishing that superconductivity does not arise from any spurious phase. The observe values of $T_c^C$, defined as the midpoint of the raising curve of C versus T, are in agreement with those obtained from the ρ(T) data as shown in figure 4. We have fitted the data in a narrow (3 to 5 K) temperature region above $T_c$ to the Debye formula, $C= \gamma T + \beta T^3$, and the values of linear coefficient (γ) of heat-capacity fall in the range 6 to 7 mJ/mol K$^2$. The values of ΔC/ γT at $T_c$ turns out to be about 1 for all cases, which is marginally lower than the weak coupling value of 1.35; for comparison, the corresponding values for CaAlSi and CaGaSi are 1.71 and 1.57 respectively [14]. The values of Debye temperature (θ$_D$) determined from the relation, θ$_D^3$ =1944$r$/β (Ref. 15) increases with $x$, if we plug a value of 6 for the number of atoms, $r$, in this relation; for instance, the value θ$_D$ turns out to be 255 K and 320 K for x= 0 and 0.7 respectively [16]. Above 5 K, the plot C/T versus T$^2$ tends to deviate from linearity and hence a marginal uncertainty may exist in the absolute values of θ$_D$. However, with this trend in θ$_D$, qualitatively speaking, following the arguments in Ref. 10, we believe that the electron-phonon coupling strength decreases with increasing Pt substitution. There is also a non-monotonic variation of the values of C(T) with *x* at any given temperature above $T_c$, and it is possible that this is due to the influence of the secondary phase.

We show the dc χ(T) data at low temperatures obtained with SQUID in the presence of a magnetic field of (notionally) 50 Oe in figure 6 for the ZFC condition of the specimens in the ingot form. The sign of χ becomes distinctly negative at the onset of superconducting transition. The results obtained with VSM on the powder form for three compositions are shown in figure 7 to compare the diamagnetic fractions for H= 20 Oe. The Meissner volume fraction estimated from the data at 2 K on powders turns out to be small, typically in the range 5-10%. This value for Pt end member is however large compared to that reported in Ref. 9. The smaller Meissner fraction is not against bulk superconductivity, considering that the heat capacity features are very prominent as described above. Reduced Meissner effect can be attributed the trapping of the magnetic flux due to large flux pinning force (for the FC condition); in addition, there is also an uncertainty in the values of magnetic field due to the difficulties in estimating and nullifying the remnant magnetic field during measurements. Finally, from a comparison of figures 6 and 7, it may be inferred that the magnetic field dependence of $T_c$ is negligible for x= 0, whereas it increases with increasing Pt content, e.g., for x= 1.0, the (onset) values for H= 50 and 20 Oe are 1.8 and 2.1 K. Future work may address this issue further.

Summarizing, the main point of emphasis in this work is that *$T_c$ in the present pseudo-ternary series varies monotonically with x unlike in all other solid solutions known till to date for this structure. This is in sharp contrast to the isoelectronic substitution even at the Ge site.* Apparently, the properties correspond to a composition close to $Y_2(Pd,Pt)_{1.3}Ge_{2.5}$. Following Ref. 11, it is possible that the composition with longer *a*-axis has a higher $T_c$ and therefore a reduction of overlap of valence orbitals in the basal plane (that is, in the Pd, Pt, Ge honeycomb layer) and a simultaneous decrease of Y-Y distance along *c*-axis with decreasing *x* change the electronic structure in favor of enhanced $T_c$. Band structure calculations [3] revealed that Y4d states dominate $E_F$ for *x*= 0.0 and we presume that there is a decrease in Y 4d electronic states due to decreasing Y4d-hybridization along c-axis with increasing *x*. Needless to point out that Si substitution for Ge reduces molecular weight, whereas Pt substitution for Pd increases the same. Despite this difference, $T_c$ decreases for both the substitutions. Therefore, the present results endorse the view [11] that the electronic structure changes, rather than electron-phonon interaction, may be the prime factor determining observed variations.

We thank Niharika Mohapatra for help during magnetization measurements employing VSM.

## References
*Corresponding author; e-mail address: sampath@tifr.res.in
1.   Jun Nagamatsu, Norimasa Nakagava, Takahiro Muranaka,Yuji Zenitani, Jun Akimitsu, *Nature (London)* 410 (2001) 63-64.
2.   Subham Majumdar, E. V. Sampathkumaran, *Phys. Rev. B* 63 (2001) 172407.
3.   E.V. Sampathkumaran, Subham Majumdar, W. Schneider, S.L. Molodtsov and C.

Laubschat, Physica B 312-313 (2002) 312.
4. M. Imai, E. Abe, J. Ye, K. Nishida, T. Kimura, K. Honma, H. Abe, and H. Kitazawa, Phys. Rev. Lett. 87 (2001) 077003; R.L. Meng, B. Lorenz, Y.S. Wang, J. Cmaidalka, Y.Y. Sun, Y.Y. Xue, J.K. Meen, and C.W. Chu, Physica C 382 (2002) 113.
5. M. Imai, K. Nishida, T. Kimura and H. Abe, Appl. Phys. Lett. 80 (2002) 1019.
6. B. Lorez, J. Lenzi, J. Camaidalka, R.L. Meng, Y.Y. Sun, Y.Y. Xue, and C.W. Chu, Physica C 383 (2002) 191.
7. H.H. Sung and W.H. Lee, Physica C 406 (2004) 15.
8. W.J. Hor, H.H. Sung, and W.H. Lee, Physica C 434 (2006) 121.
9. H. Kito, Y. Takano and K. Togano, Physica C 377 (2002) 185; these authors report a higher value of 3.2 K for $T_c$ of $Y_2PtGe_3$.
10. A.K. Ghosh, H. Nakamura, M. Tokunaga, and T. Tamegai, Physica C 388-389 (2003) 567.
11. S. Takano, Y. Iriyama, Y. Kimishima, and M. Uehara, Physica C 383 (2003) 295.
12. See, for example, F.A. Schmidt, O.D. Mc Master, and O.N. Carlson, J. Alloys and Compounds, J. Alloys and Comp. 26 (1972) 53.
13. See, Landoldt Bernstein Numerical Data and Functional Relationships in Science and Technology, New Series III, Editor-in-Chief: O. Madelung, Vol 21a, page 283.
14. T. Tamegai, K. Uozato, S. Kasahara, T. Nakagawa, and M. Tokunaga, Physica C 426-431 (2005) 208.
15. See, "Specific heats at low temperatures" by E.S.R. Gopal (Plenum, New York) 1966.
16. In the previous reports (Ref. 2 and 10), the value of $\theta_D$ for x= 0.0 was nearly half of present value though the heat-capacity values reported here are similar to earlier values. The origin of this error in the earlier reports is not clear.

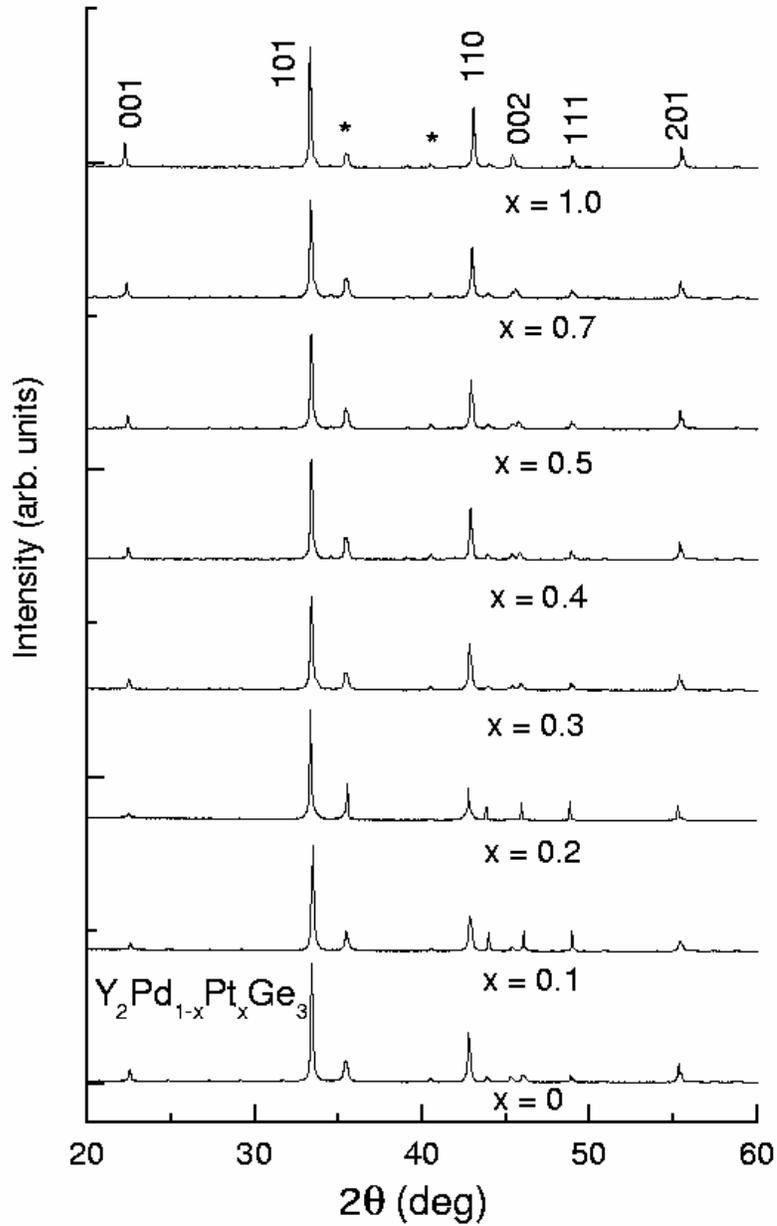

Figure 1:
(color online) X-ray diffraction patterns (Cu $K_\alpha$) for $Y_2Pd_{1-x}Pt_xGe_3$ alloys. Asterisks represent parasitic $ThCr_2Si_2$-type tetragonal phase. The curves are physically shifted along y-axis for the sake of clarity.

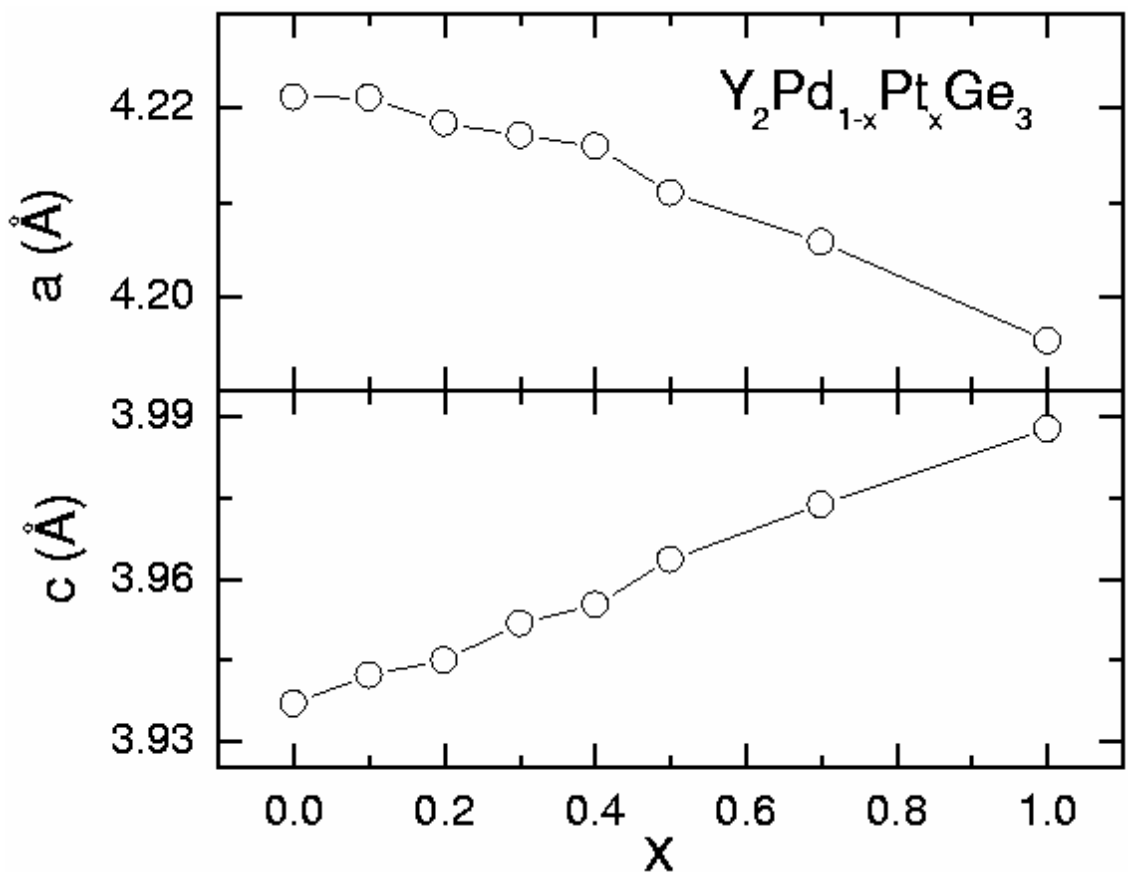

Figure 2:
The lattice constants, *a* and *c*, as a function of *x* in the solid solution, $Y_2Pd_{1-x}Pt_xGe_3$. The lines through the data points serve as a guide to the eyes.

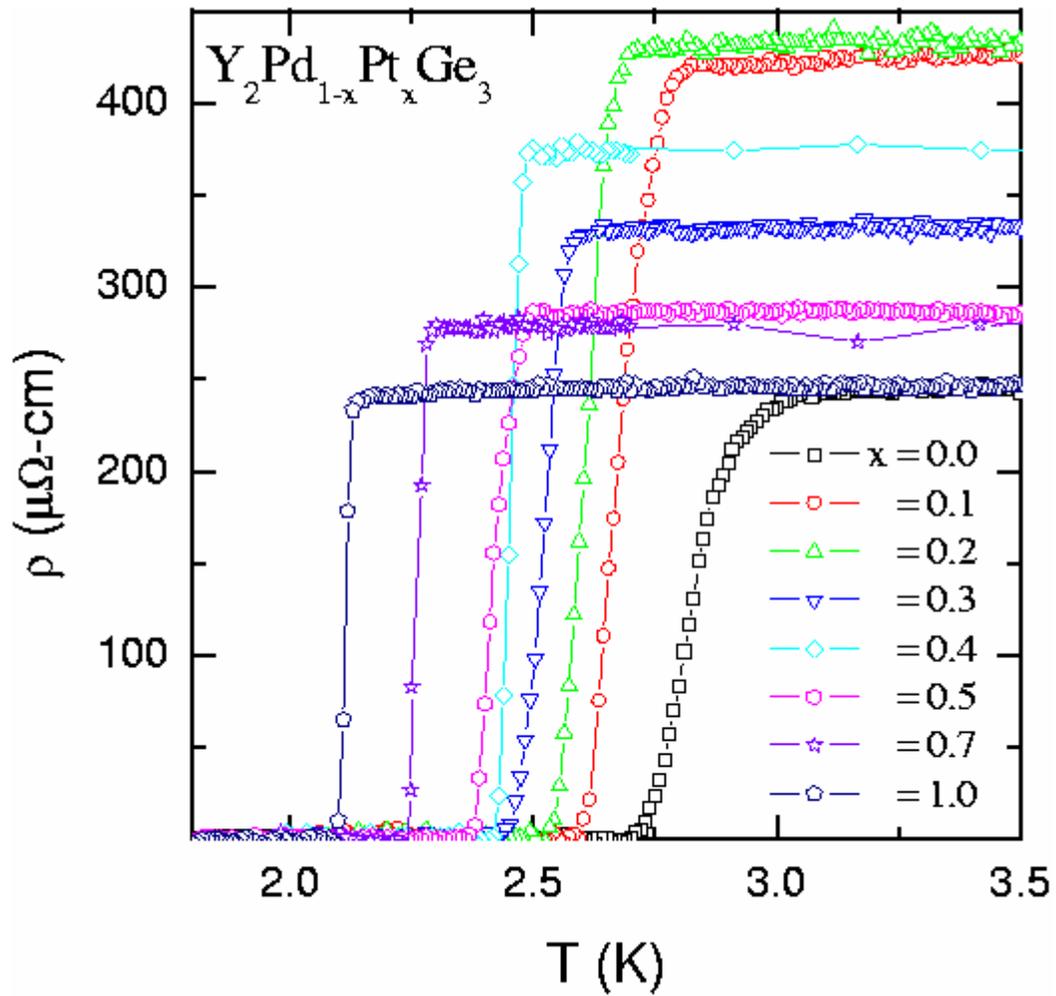

Figure 3:
(color online) Electrical resistivity as a function of temperature below 3.5 K for the alloys, $Y_2Pd_{1-x}Pt_xGe_3$. The lines through the data points serve as a guide to the eyes.

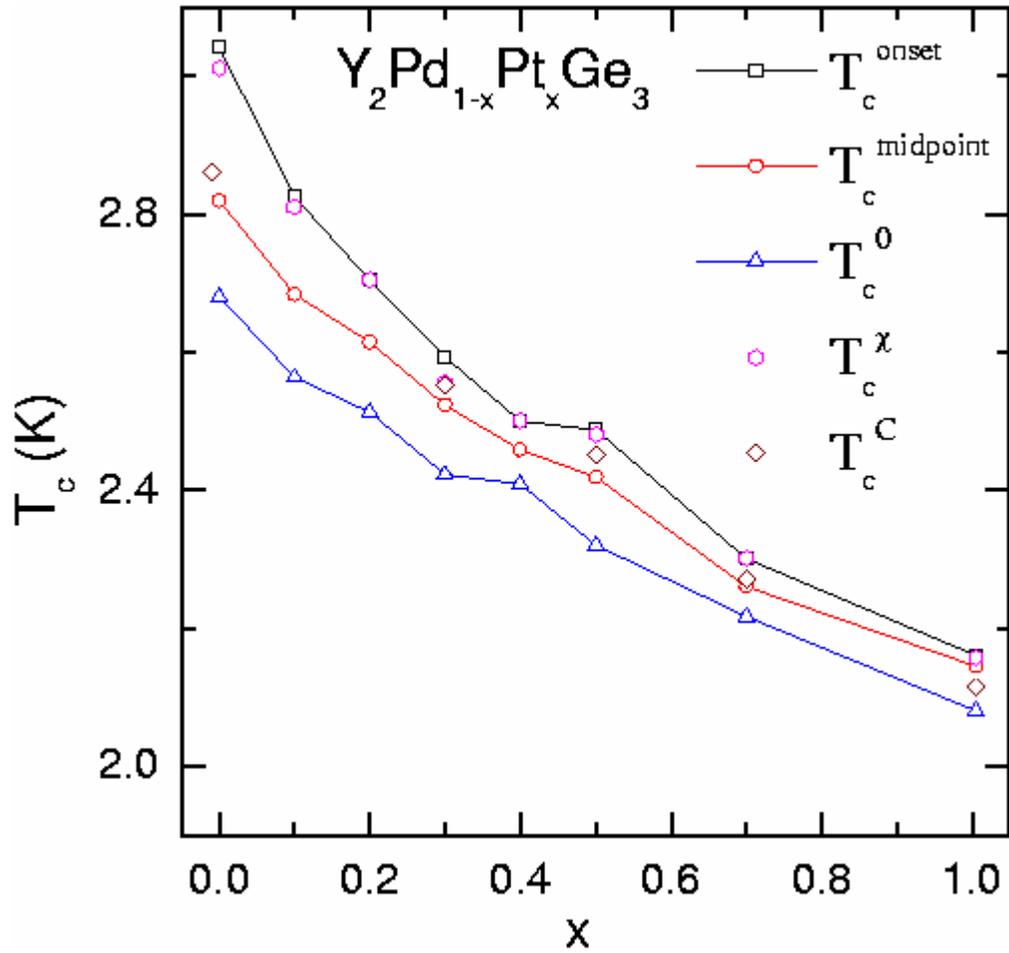

Figure 4:
(color online) $T_c^{onset}$, $T_c^{midpoint}$, and $T_c^0$ for the alloys, $Y_2Pd_{1-x}Pt_xGe_3$, as determined from the data shown in figure 3. The lines through the data points serve as a guide to the eyes. The transition temperatures, $T_c^C$ and $T_c^\chi$, inferred from C(T) and χ(T) data, are also shown.

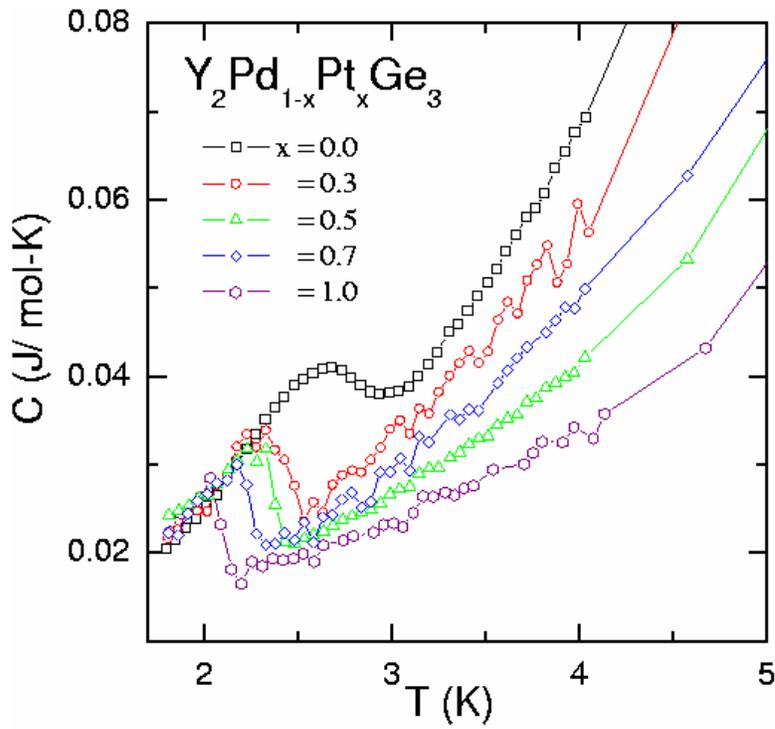

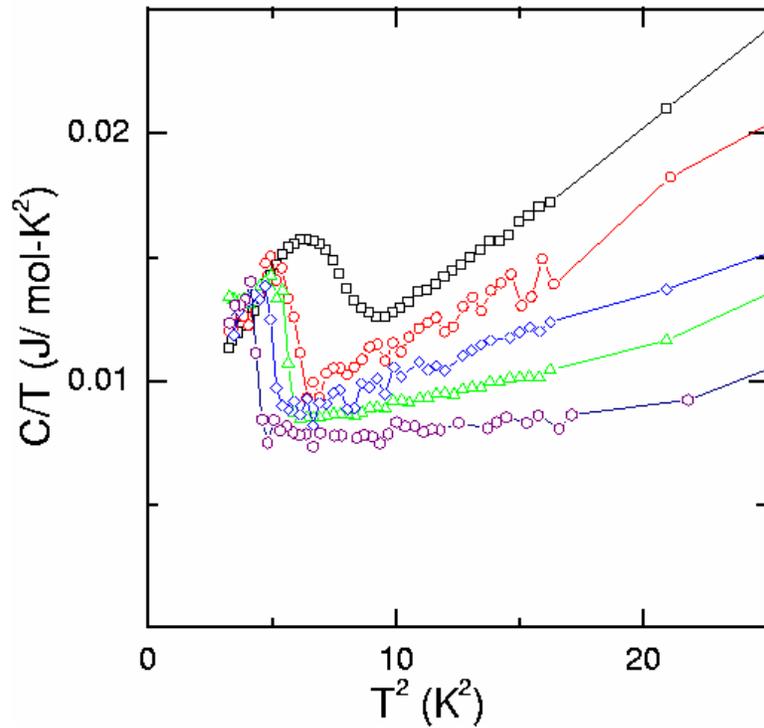

Figure 5:
(color online) Heat capacity (C) as a function of temperature (T) and C/T versus $T^2$ below 5 K for the alloys, $Y_2Pd_{1-x}Pt_xGe_3$. The lines through the data points serve as a guide to the eyes.

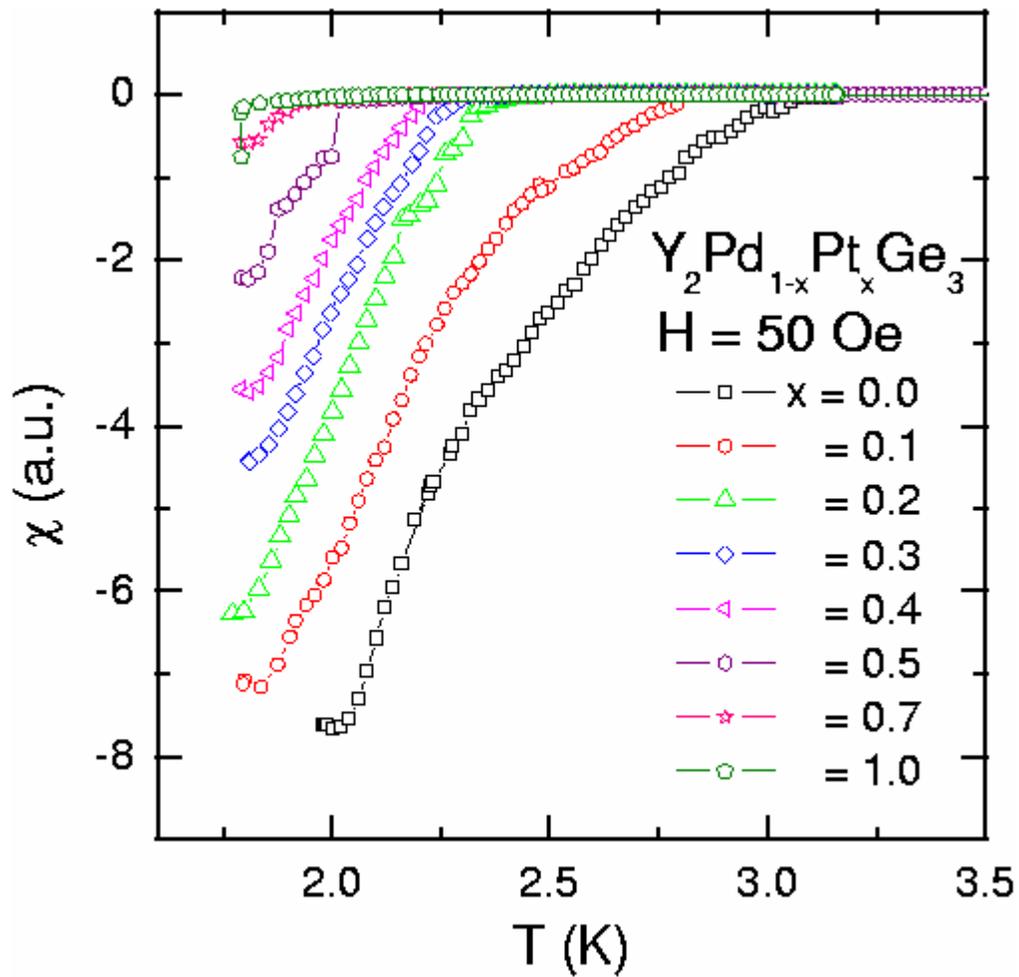

Figure 6:
(color online) Magnetic susceptibility behavior at low temperatures obtained in a field of 50 Oe for the alloys, $Y_2Pd_{1-x}Pt_xGe_3$ for the zero-field-cooled conditions of the specimens, obtained with a SQUID magnetometer. The lines through the data points serve as a guide to the eyes.

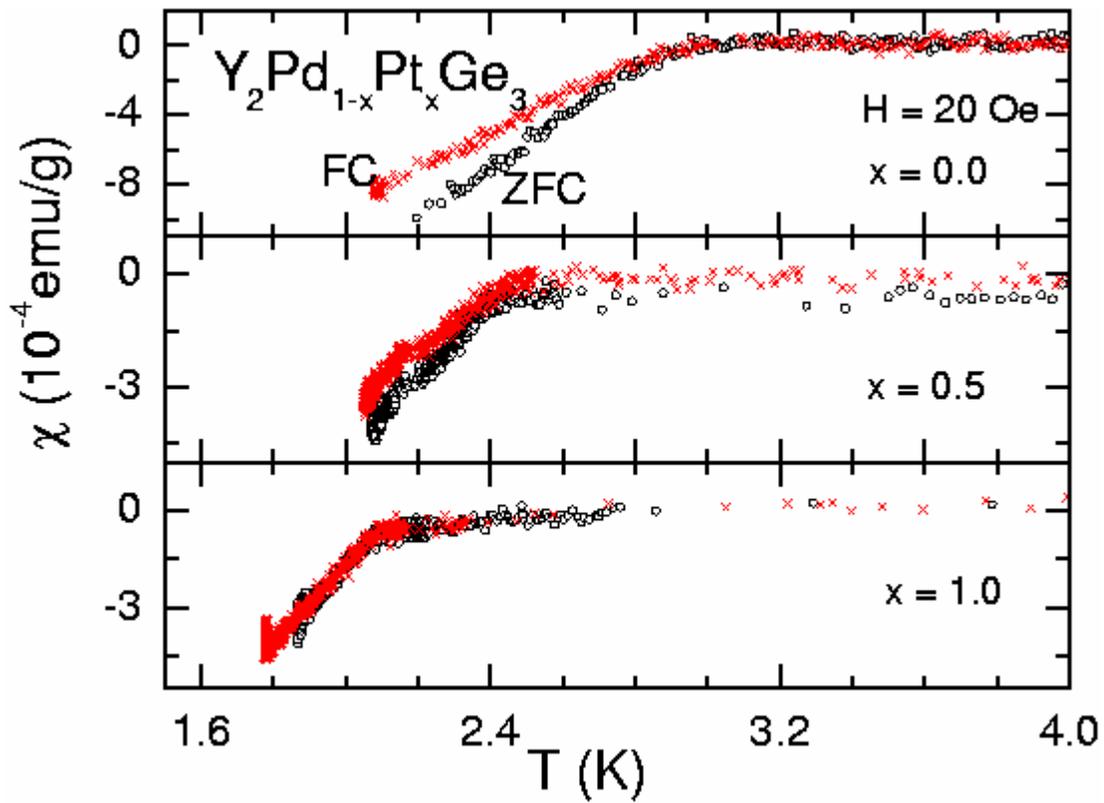

Figure 7:
(color online) Magnetic susceptibility behavior at low temperatures obtained in a field of 20 Oe for the alloys, $Y_2Pd_{1-x}Pt_xGe_3$ (x= 0.0, 0.5 and 1.0) for zero-field-cooled and field-cooled conditions of the specimens, obtained with a VSM magnetometer.